\documentclass{PoS}

\title{Four and two-lepton signals of leptophilic gauge interactions at large colliders}

\ShortTitle{Leptophilic interactions at large colliders}

\author{\speaker{Francisco del Aguila} \thanks{We are grateful to the Corfu Summer Institute 2014 organizers 
for their hospitality.} $\ ^a$, 
Mikael Chala $^b$, Jos\'e Santiago $^{a, c}$ and Yasuhiro Yamamoto $^a$
\vspace{0.3cm} \\
$^a$ CAFPE and Departamento de F{\'\i}sica Te{\'o}rica y del Cosmos, Universidad
de Granada, E\textendash{}18071 Granada, Spain 
\vspace{0.1cm} \\
$^b$ DESY, Notkestrasse 85, 22607 Hamburg, Germany 
\vspace{0.1cm} \\
$^c$ Institute for Theoretical Physics, ETH Z\"urich, 8093 Z\"urich, Switzerland
\vspace{0.3cm} \\
E-mail: \email{faguila@ugr.es}, \email{mikael.chala@desy.de}, \email{jsantiago@ugr.es}, \email{yamayasu@ugr.es} 
\vspace{0.3cm} \\}

\abstract{Many Standard Model extensions can contribute to four-lepton signals at large colliders. 
We review the particular case of leptophilic interactions eventually observable at the LHC and the ILC, 
paying special attention to the addition of a new vector boson coupled to muon minus tau lepton number, 
$Z'_{\mu - \tau}$, and emphasizing the prospects at a very large hadron collider with $\sqrt s =$ 100 TeV. 
We also discuss in this case the new contribution to two-lepton (Drell-Yan) production when the new leptophilic 
interaction has a non-vanishing kinetic mixing with the SM.
}

\FullConference{Proceedings of the Corfu Summer Institute 2014 \\
		 3-21 September 2014\\
		 Corfu, Greece}

\begin{document}

\section{Introduction}

The Standard Model (SM) has been completed with the discovery of the 
Brout-Englert-Higgs (BEH) boson \cite{Englert:1964et} at the Large Hadron 
Collider (LHC) \cite{Aad:2012tfa}. 
Besides, the outstanding performance of the machine and of the LHC 
experimental collaborations have also allowed to translate the non-observation 
of any significative departure from the SM predictions into stringent limits on 
New Physics (NP). 
What, in particular, implies that if NP is going to show up at future runs, it must 
predict relatively small cross-sections into final states already scrutinized. 
Obviously, small cross-sections can be always obtained banishing the NP scale 
to higher energies or coupling it to the SM weakly enough. 
The relevant question is if there is room left for new interactions near the TeV 
scale with effective couplings of at least Electro-Weak (EW) strength.   
What also brings to the question of whether NP can manifest at the 
International Linear Collider (ILC) if no departure from the SM can be established 
at the LHC.

Specific SM extensions can predict some new particles to be in practice invisible 
at the LHC even if their masses are near the EW scale. Indeed, the LHC discovery 
reach can be reduced allowing these heavy particles to decay into new channels 
with larger backgrounds, making in any case them wider and eventually 
undetectable.   
Their production, and then their detection, can be also made more difficult 
if they are constrained to be pair-produced by imposing the corresponding selection 
rule. 
There are many of these examples in the literature. For instance, 
stealth gluons \cite{Barcelo:2011vk} 
provide an example of the first case, and supersymmetric models 
with R-parity \cite{Farrar:1978xj} of the second one. 
Often these properties reducing the LHC discovery potential are also retained at the ILC, 
not being then this $e^+e^-$ machine an alternative for these NP searches. 
Although the actual sensitivity for a given 
process will depend on which circumstance plays a more important role, 
its lower Center of Mass Energy (CME) or its cleaner environment, and 
eventually the available beam polarization at this leptonic machine \cite{Baer:2013cma}. 

\begin{figure}
\begin{centering}
\includegraphics[width=0.49\columnwidth]{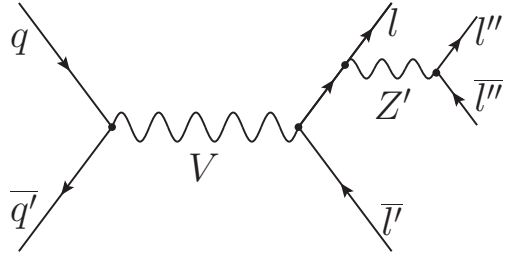}
\par\end{centering}
\caption{Leading diagram for $Z'_{\mu - \tau}$ production at the LHC, for negligible kinetic mixing.
\label{Productiondiagram}
}
\end{figure}

One can be more drastic, however, and assume that the new particles 
do not couple to quarks and gluons, nor to EW bosons, but only 
to SM leptons, at least to start with. Trying in this way to reduce as much 
as possible the LHC discovery reach, but not the ILC one. 
Nonetheless, even in this extreme scenario the LHC seems to be able 
of partially settling the question, whenever electrons (positrons) are not relevant. 
A detailed study of the LHC and the ILC potential for the discovery of these leptophilic 
interactions is presented in Ref. \cite{delAguila:2014soa}. 
We summarize the main results presented there in Section \ref{Colliderlimits}, 
but concentrating on the particular case of an extra gauge boson coupling to muon minus tau Lepton Number (LN), 
$Z'_{\mu - \tau}$ \cite{Foot:1994vd}, as example. 
We then compare the LHC and ILC potentials with the potential of a Future 
Circular hadron Collider (FCC) with a CME $\sqrt s = 100$ TeV 
\cite{hadronFCC} 
\footnote{Future Circular lepton Colliders have also deserved consideration \cite{leptonFCC}, 
as well as a Large Hadron electron Collider (LHeC) \cite{AbelleiraFernandez:2012cc} with 
different options.} 
in Section \ref{FCClimits}. 
Although such a leptophilic gauge boson tends to mix little with the SM $Z$ boson, the kinetic mixing 
can be large in generic models \cite{Babu:1997st} and then, the $Z'_{\mu - \tau}$ 
contribution to the Drell-Yan process, $q\bar q \rightarrow Z'_{\mu - \tau} \rightarrow l\bar l$, 
could be eventually sizable at a large hadron collider. 
We discuss this scenario in Section \ref{Gaugemixing}, finding that the limits on 
the complementary process at the LEP 2, $e^+ e^- \rightarrow Z'_{\mu - \tau} \rightarrow l\bar l$, 
already constrain the kinetic mixing to be small for relatively low vector 
boson masses of few hundreds of GeV. 
However, Drell-Yan production at the LHC (FCC) 
provides the most stringent bound on such a mixing, 
especially for larger masses of the order of the TeV. 
Thus, four-lepton production in Fig. \ref{Productiondiagram} is the only leptophilic 
final state available for vanishing kinetic mixing but $q\bar q \rightarrow \gamma, Z, Z'_{\mu - \tau} \rightarrow l\bar l$ 
allows to set quite strong limits on these interactions for large kinetic mixing.  
Final comments are collected in the Conclusions.

\section{Limits on $Z'_{\mu - \tau}$ at the LHC and the ILC}
\label{Colliderlimits}

As we have recently argued in Ref. \cite{delAguila:2014soa}, the only leptophilic particle 
coupling to SM lepton pairs in a renormalizable way is a new neutral vector boson $Z'$. 
If this boson was a scalar, it would have to be the component of a multiplet transforming as the BEH one or including a 
doubly-charged component, too (see Ref. \cite{delAguila:2013yaa} for a detailed discussion of the lowest 
order couplings and production mechanisms for such a scalar). 
Thus, implying in either case that it would couple to SM gauge bosons with 
EW strength and hence, it would not be a leptophilic particle. 
(Although at the end, leptophilic additions are expected to partially lose their character 
because higher order effects may result in a small mixing between the different 
sectors of the theory, thus coupling to EW gauge bosons and to quarks, too, but 
with suppressed couplings.) 
Bosons with spin higher than 1 do not have renormalizable couplings to lepton pairs, either.
The generic Lagrangian describing the corresponding vector boson interactions reads:  
\begin{equation}
{\cal L}_{Z'} = - (g^{\prime ij}_{\rm L} {\overline L_{i {\rm L}}} \gamma^\mu L_{j {\rm L}}  + 
g^{\prime ij}_{\rm R} {\overline l_{i {\rm R}}} \gamma^\mu l_{j {\rm R}}) Z'_\mu \ ,
\label{zprimecouplings}
\end{equation}
where $g^{\prime ij}_{{\rm L, R}}$ are arbitrary dimensionless couplings to 
the SM Left-Handed (LH) and Right-Handed (RH) lepton multiplets, 
$L_{i {\rm L}} = \left( \begin{array}{l} \nu _{i {\rm L}} \\ l_{i {\rm L}} \end{array} \right)$ 
and $l_{i {\rm R}}$, respectively, with $i = e, \mu, \tau$ labeling the charged-lepton 
family. 
Even though we allow for arbitrary flavor and chiral interactions, 
the EW gauge symmetry remains unbroken because the new couplings of the LH charged 
leptons and their neutrino counterparts are equal. However, the Yukawa couplings giving masses 
to the charged leptons in the SM do not need to preserve such a hypothetical gauge symmetry. 
As a matter of fact, if a leptonic Yukawa term has to be invariant under the new symmetry, 
the scalar field coupling to the pair of LH and RH leptons must have a 
commensurate charge adding to zero with the lepton charges. 
Thus, in general, the existence of a new leptophilic interaction described 
by Eq. (\ref{zprimecouplings}) requires non-minimal scenarios with a non-standard 
lepton mass generation and/or extra scalar multiplets for the model to be realistic. 

An example of the first scenario, without extra light scalar multiplets and hence, without Yukawa 
terms involving leptons (because the SM BEH doublet must be a singlet under the 
new symmetry in order to quarks get SM masses), is the composite Higgs model with 
tau custodians in Ref. \cite{delAguila:2010vg}. In this case, once the extra heavy 
vector bosons present in the model acquire a mass, lepton Yukawa couplings are 
generated at higher orders in perturbation theory.   

A prime example of a model with an extra leptophilic interaction and possibly extra scalar doublets 
is the combination of muon minus tau LN, ${\rm L}_\mu - {\rm L}_\tau$,  
\footnote{In order to give a large enough mass to the new gauge boson, 
the model must also include at least one extra scalar field 
transforming trivially under the SM gauge symmetry but not under ${\rm L}_\mu - {\rm L}_\tau$.} 
with diagonal couplings $g^{\prime \mu\mu, \tau\tau}_{{\rm L,R}}$ given by the charges  
\begin{equation}
\begin{tabular}{l|cccc}
Multiplet \quad & 
\quad $L_{{\rm L} \mu} = \left( \begin{array}{c}  \nu_{\mu {\rm L}} \\ \mu_{\rm L} \end{array} \right)$ \quad & 
\quad $\mu_{\rm R}$ \quad & 
\quad $L_{{\rm L} \tau} = \left( \begin{array}{c}  \nu_{\tau {\rm L}} \\ \tau_{\rm L} \end{array} \right)$ \quad & 
\quad $\tau_{\rm R}$ \quad  \\
\hline
Charge & 
$1$ & 
$1$ & 
$- 1$ & 
$- 1$
\end{tabular}
\end{equation}
times the gauge coupling constant $g'$. 
This is the only gauge addition to the SM with vanishing couplings to electrons 
and anomaly free in the absence of extra fermions \cite{Foot:1994vd}. 
On the other hand, this model also allows for a good estimate (which can be also easily extrapolated \cite{delAguila:2014soa}) 
of the LHC (and the ILC) potential for the discovery of leptophilic interactions. 
Its reach appears to be relatively reduced (in either case) because the two-lepton signal of the new resonating 
vector boson must in principle emerge from a four-lepton sample. This means that we must deal 
with relatively small cross-sections of few fb at the LHC for $\sqrt s = 14$ TeV (and similarly 
at the ILC) and hence, that there is only enough sensitivity for an eventual discovery 
when the new gauge couplings are relatively large and/or the $Z'$ masses are somewhat small. 
This then excludes possible electron signals if the limits from LEP 2 must be fulfilled \cite{Schael:2013ita}. 
Assuming universality, for these limits do not show a large flavor dependence, 
the corresponding 95 \% Confidence Level (C.L.) bounds read 
\cite{Blas:2013ana} (see also \cite{delAguila:2010mx}):  
\begin{equation}
\frac{g'^{ee}_{\rm L, R}}{M_{Z'}} < 0.12, 0.16 \ {\rm TeV}^{-1} \ ,
\label{zprimebounds}
\end{equation}
being too stringent for observing a leptophilic vector boson in a four-lepton final state at the LHC (ILC). 
We assume that the model gauging ${\rm L}_\mu - {\rm L}_\tau$ also includes a SM singlet scalar 
giving a relatively large mass to the new gauge boson. 
The originally proposed realization addressed the possibility of a relatively light $Z'_{\mu - \tau}$. 
In our case, the extra vector boson can be made arbitrarily heavy. We will advocate for 
the extra scalar doublets transforming non-trivially under the new gauge symmetry 
only to generate an appreciably mass mixing with the SM gauge bosons, when needed. 
Neutrino masses are very tiny 
and we assumed they are generated at higher orders in a non-minimal scenario. 
In any case, in the following the parameters used to fix the model in the Monte Carlo 
simulations shall be only the new gauge coupling $g'$ and the heavy vector boson mass 
$M_{Z'_{\mu - \tau}}$; while the total width writes: 
\begin{equation}
\Gamma_{Z'_{\mu - \tau}} = \frac{g'^2}{4 \pi} M_{Z'_{\mu - \tau}} \ ,
\label{width}
\end{equation}
where we neglect any $Z'_{\mu - \tau}$ decay into new scalar pairs. 

\begin{figure}
\begin{centering}
\includegraphics[width=0.49\columnwidth]{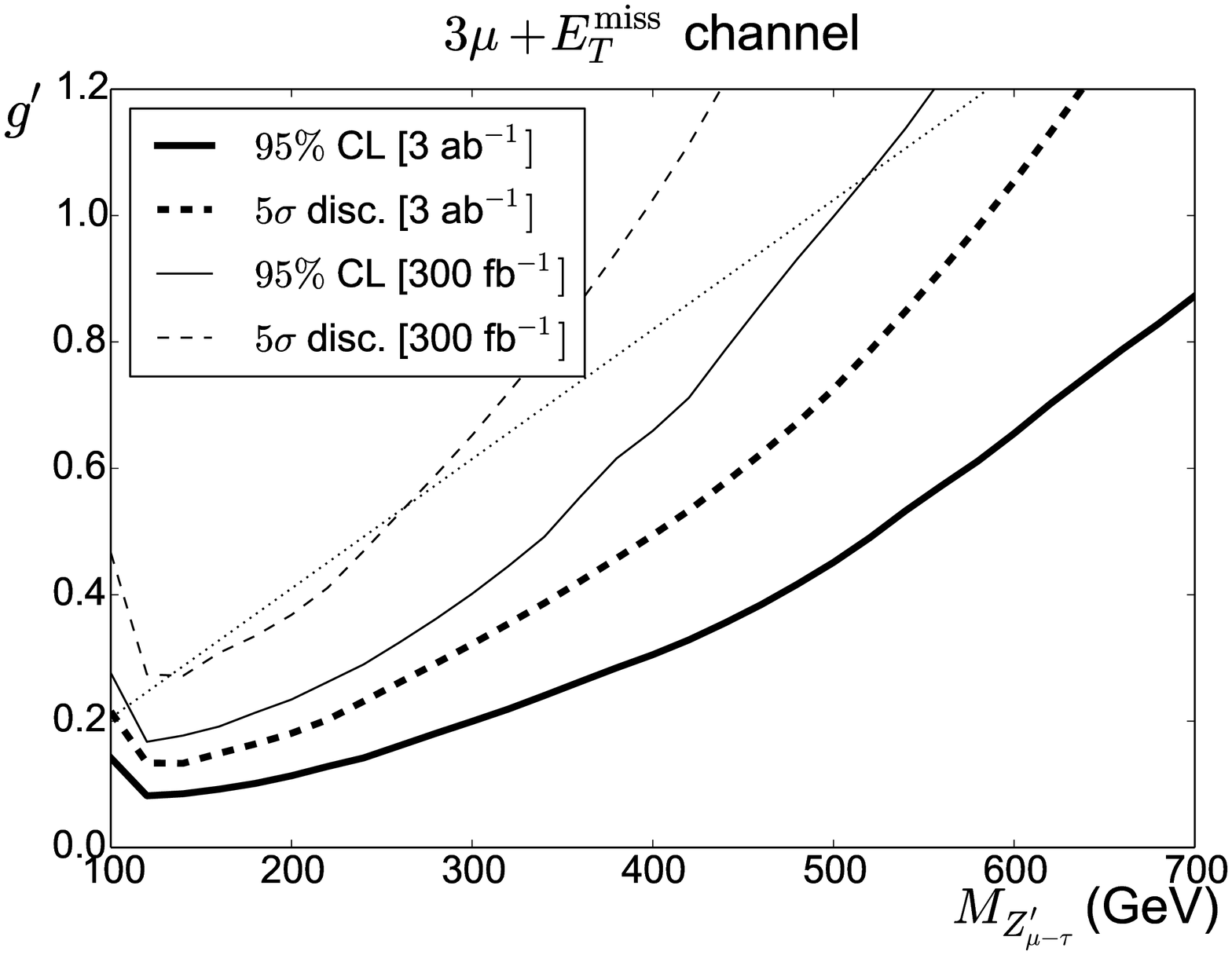}
\includegraphics[width=0.49\columnwidth]{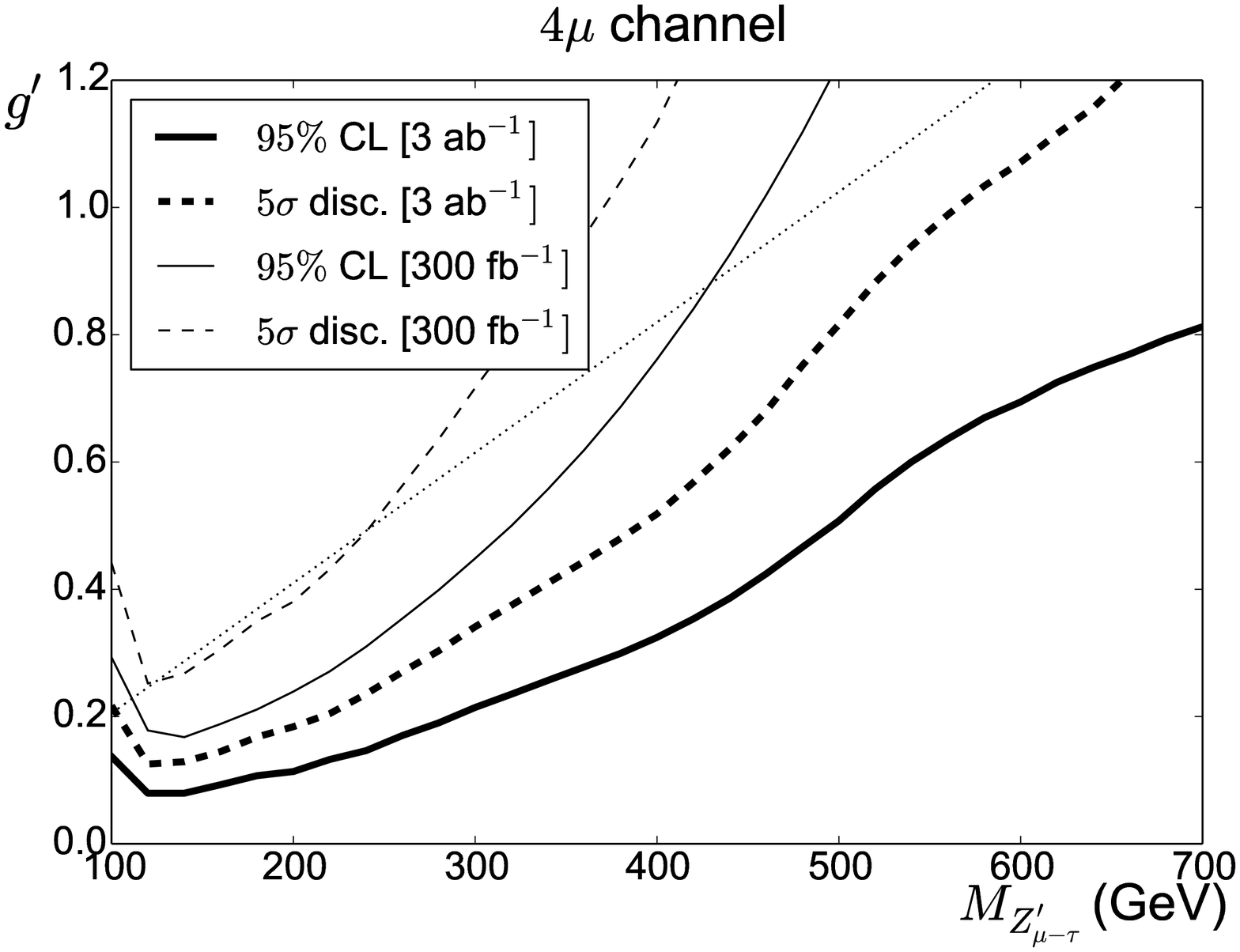}
\par\end{centering}
\caption{Discovery and exclusion limits for $Z'_{\mu - \tau}$ as a function 
of its mass at the LHC with $\sqrt s = 14$ TeV. On the left (right) we plot the bounds obtained 
from 3$\mu$ plus missing energy (4$\mu$) events. 
We also draw the bounds from neutrino trident production (straight line), %and $g-2$, 
for comparison (see the text for details).
\label{Mgdiscovery}
}
\end{figure}

The LHC and ILC limits on a generic leptophilic vector boson only coupling to muons 
and taus are throughly discussed in Ref. \cite{delAguila:2014soa}. 
Let us review here the case of the $Z'_{\mu - \tau}$ above. 
In general, such a vector boson will show up in the four-lepton final states. 
As a matter of fact, the best limits at the LHC can be generally derived from the 3$\mu$ 
plus missing energy events. Although for this model the bounds from 4$\mu$ 
are somewhat similar. 
In Fig. \ref{Mgdiscovery} we plot the corresponding discovery and exclusion 
bounds for the 3$\mu$ plus missing energy (left) and 4$\mu$ (right) samples. 
As explained in Ref. \cite{delAguila:2014soa}, these curves can be 
obtained from the plots in this reference multiplying by the appropriate 
factors. In particular, the exclusion bound can be read from Fig. 3 with $\xi \equiv g'_{\rm R} / g'_{\rm L} = 1$ 
there multiplying the corresponding coupling constant for a given $Z'$ mass by $\sqrt 2$, 
to take into account the effect of the extra vector boson decay channel into taus. 
The event generation (codes used) and selection (applied cuts) are described in 
that reference, too. 
For comparison, we also show the bound from the neutrino trident production 
(straight line) \cite{Altmannshofer:2014cfa}, 
taking into account CHARM-II and CCFR data. 

\begin{figure}
\begin{center}
\includegraphics[width=0.49\columnwidth]{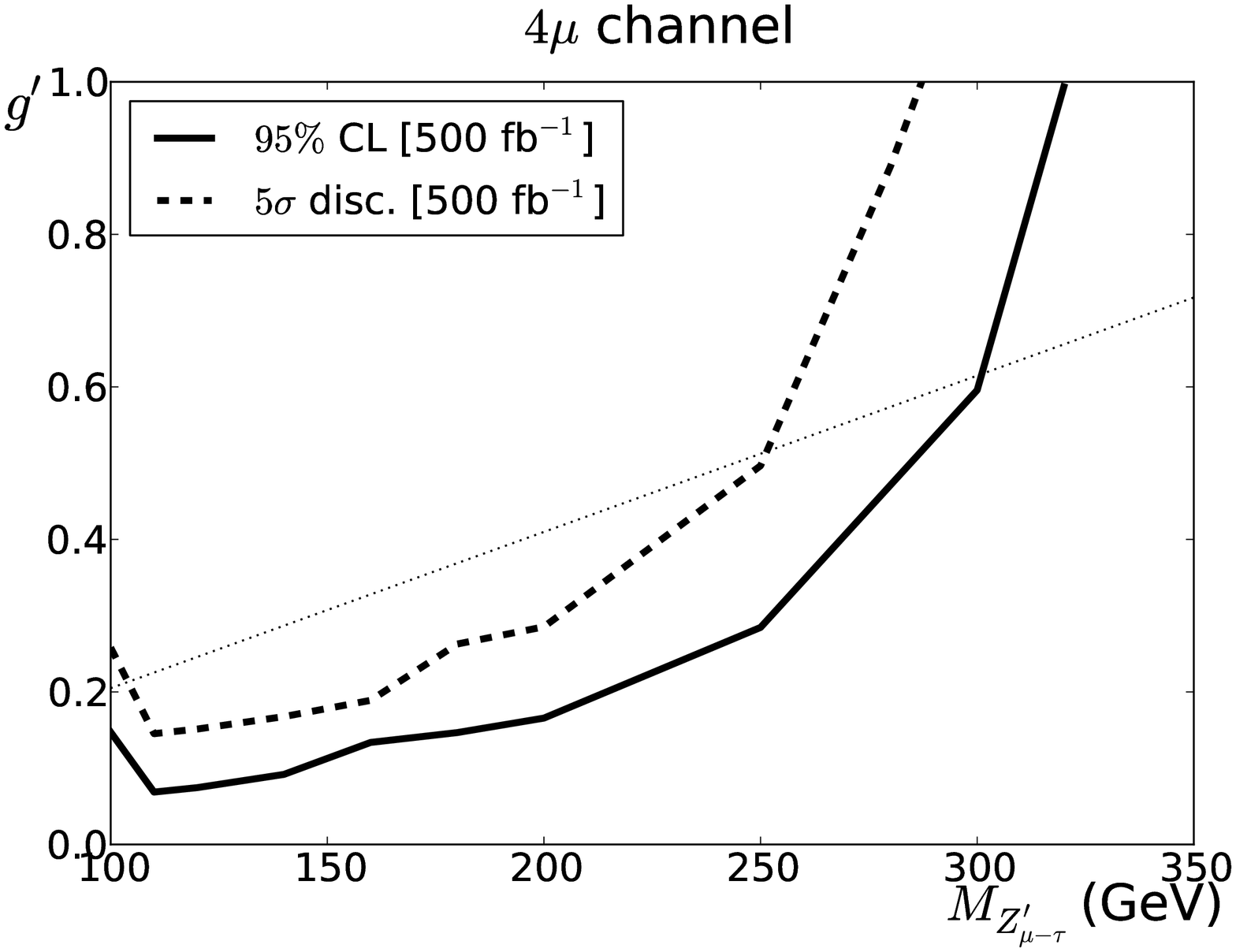}
\includegraphics[width=0.49\columnwidth]{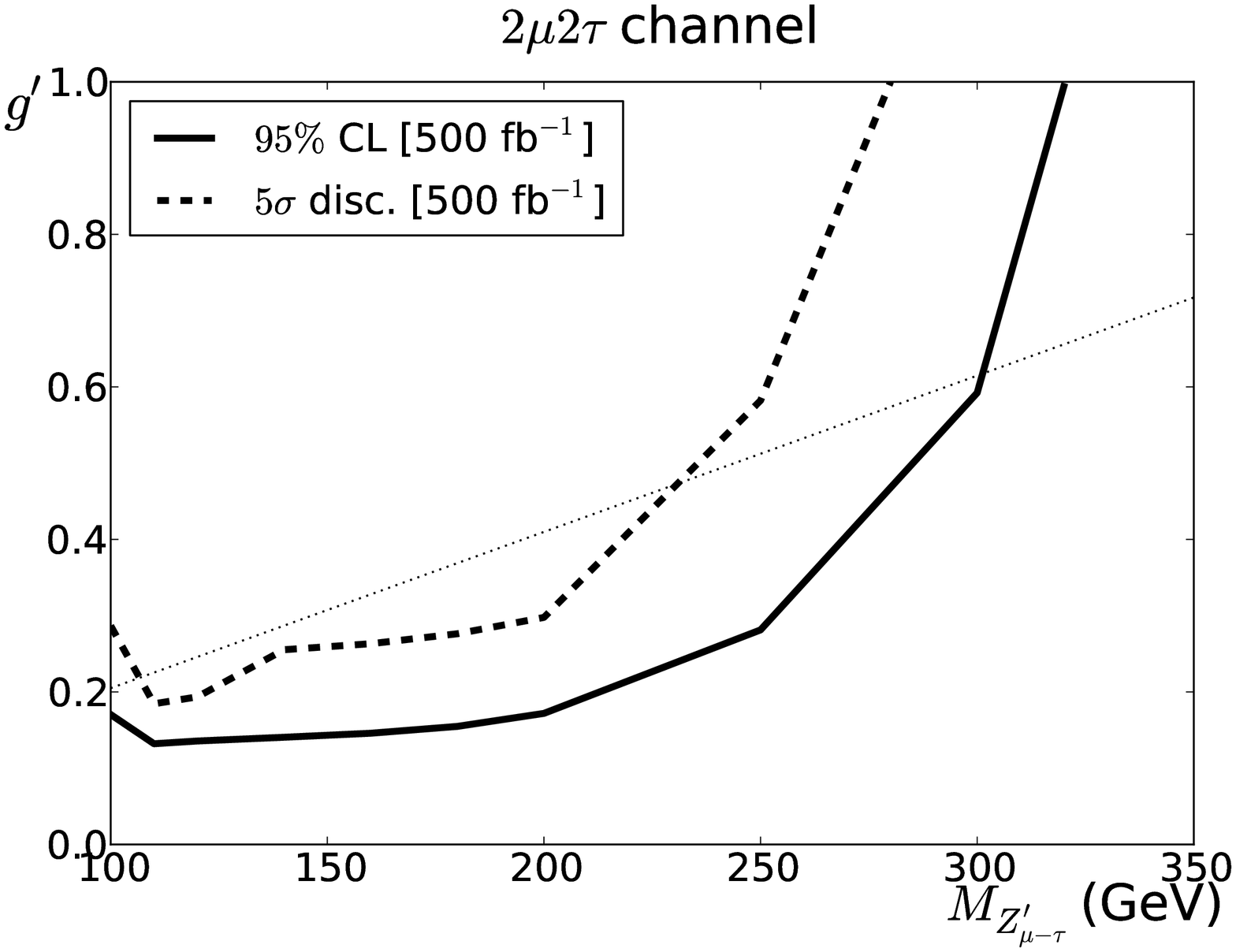}
\par\end{center}
\caption{Discovery and exclusion limits for $Z'_{\mu - \tau}$ as a function 
of its mass at the ILC with $\sqrt s = 500$ GeV. On the left (right) we plot the bounds obtained 
from 4$\mu$ (2$\mu$2$\tau$) events. 
The constraint from neutrino trident production (straight line) 
is also shown for comparison.
\label{ILCMgxi}     
}
\end{figure}

Whereas only signals involving muons can emerge at the LHC for the 
$Z'_{\mu - \tau}$ cross-sections at hand, as pointed out in Ref. \cite{delAguila:2014soa}, 
the ILC could observe the mixed channels with tau leptons. 
Thus, in Fig. \ref{ILCMgxi} we show the corresponding discovery and exclusion limits 
but for 4$\mu$ and 2$\mu$2$\tau$ events, respectively. 
Again, the bounds can be read from that reference. In particular, 
the limits obtained from 4$\mu$ events can be read from Fig. 7 (right) there 
multiplying the corresponding coupling constant for a given $Z'$ mass by $\sqrt 2$, 
to account for the extra decay channel into taus. 
Details on the event generation and selection cuts are given in the same reference. 

As emphasized in Ref. \cite{delAguila:2014soa}, 
the LHC can be sensitive to a $Z'_{\mu - \tau}$ with a mass up to $\sim 1$ TeV for 
$g'$ of order 1; while the ILC is sensitive to $Z'_{\mu - \tau}$ masses below 
$300$ GeV but it can eventually allow for a direct measurement of the 
couplings to tau leptons.

\section{Limits on $Z'_{\mu - \tau}$ at a future hadron collider with $\sqrt s = 100$ TeV}
\label{FCClimits}

\begin{figure}
\begin{centering}
\includegraphics[width=0.49\columnwidth]{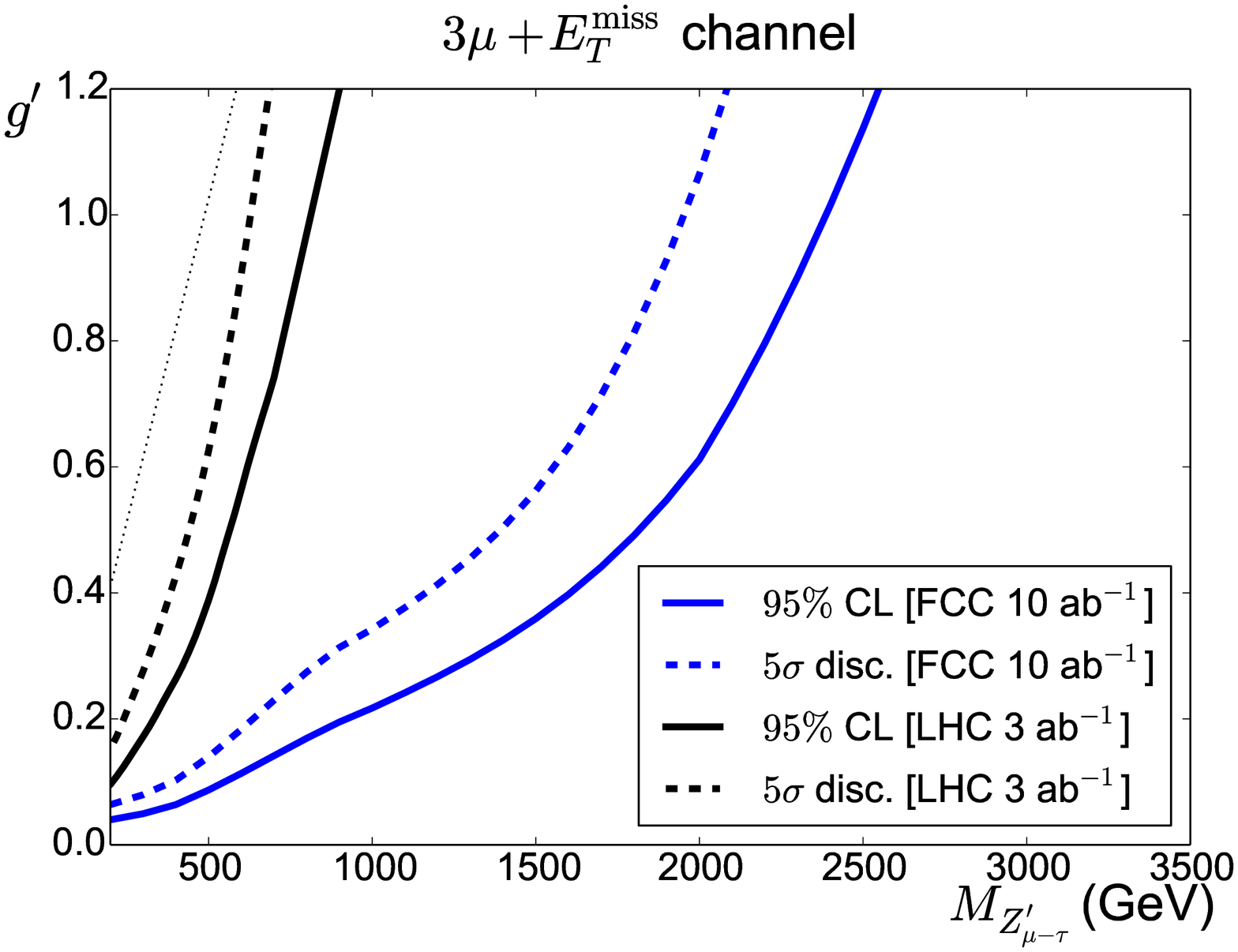}
\includegraphics[width=0.49\columnwidth]{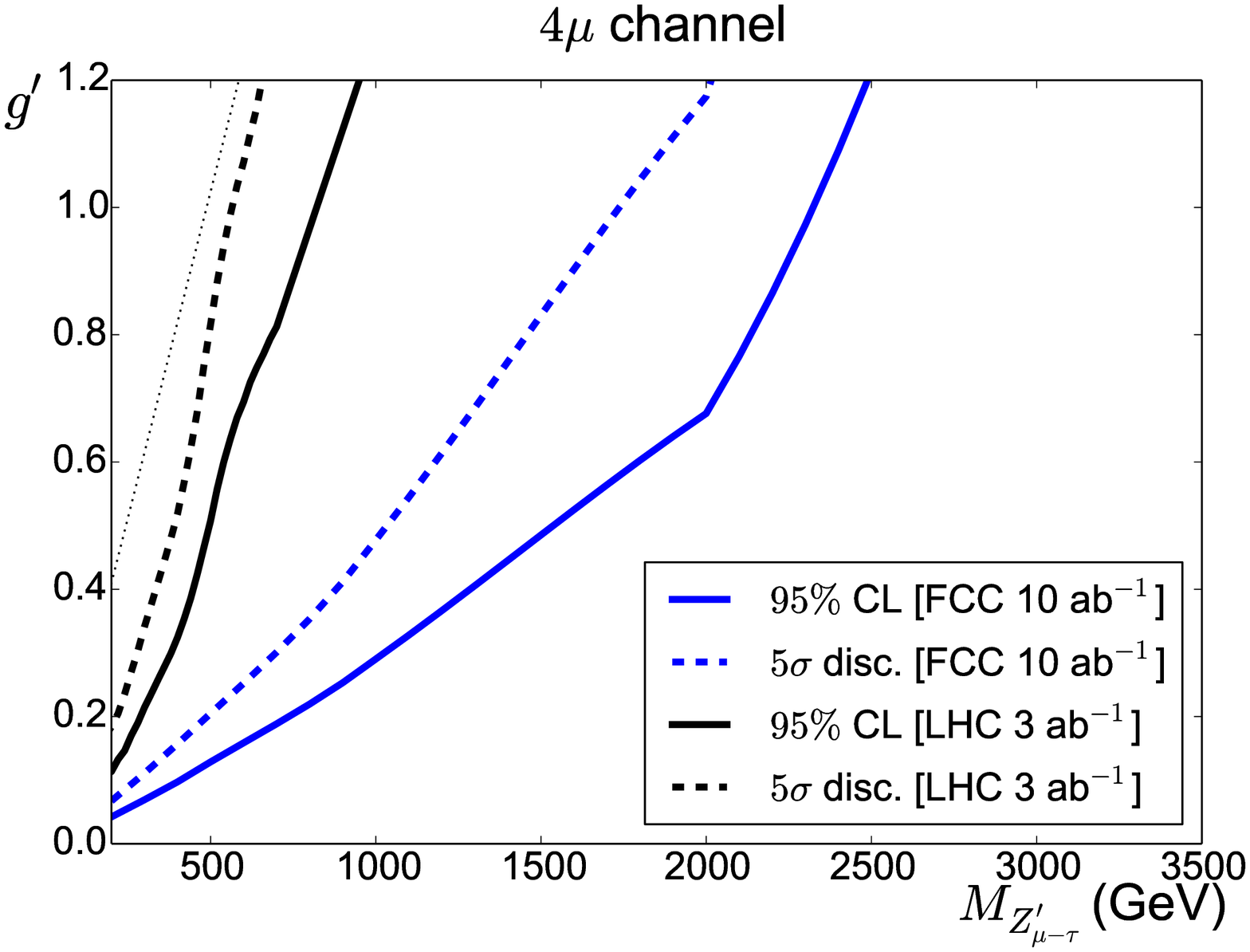}
\par\end{centering}
\caption{Discovery and exclusion limits for $Z'_{\mu - \tau}$ as a function 
of its mass at the future hadron collider (FCC) with $\sqrt s =$ 100 TeV (thick curves). 
On the left (right) we plot the bounds obtained from 3$\mu$ plus missing energy (4$\mu$) events. 
We also draw present bounds from neutrino trident production (straight line) 
and the LHC limits for $\sqrt s =$ 14 TeV (thin curves), for comparison.
\label{FCCMgdiscovery}
}
\end{figure}

As already pointed out, the LHC and the ILC will improve the limits on a new leptophilic $Z'$ 
unmixed with the SM, but for relatively large coupling constants and relatively small vector boson masses. Their reach will 
be appreciably enlarged at a larger hadron collider with a CME of $\sqrt s =$ 100 TeV, however. 
Similarly as for the LHC in the previous section, in Fig. \ref{FCCMgdiscovery} we plot the 
discovery and exclusion bounds for the 3$\mu$ plus missing energy (left) and 4$\mu$ (right) samples. 
The codes used for the event generation are the same as in Ref. \cite{delAguila:2014soa}. 
The selection cuts applied to the 3$\mu$ plus missing energy sample must be optimized, 
however, to obtain similar limits to those derived from the 4$\mu$ channel. 
This is necessary to get rid of the background events in that case, too.  
Thus, we increase the basic cut on the muon transverse momentum in Table 1 in the previous reference 
from 50 to 100 GeV; and the cuts on the missing transverse energy and on the transverse mass from 
100 to 150 GeV and from 110 to 140 GeV, respectively. 
As a result, the FCC bounds on the gauge coupling improve by almost 
a factor of 5 those at the LHC for low $Z'_{\mu - \tau}$ masses; whereas its mass reach is near 2.5 TeV for 
$g' \sim 1$.

\section{$Z'_{\mu - \tau}$ kinetic mixing with the SM and di-lepton versus four-lepton signals of a leptophilic vector boson}
\label{Gaugemixing}

There is a wide literature on the implications of the kinetic mixing between abelian gauge factors 
\cite{Holdom:1985ag} in SM gauge extensions \cite{del Aguila:1985cb}. 
We will follow Ref. 
\cite{Babu:1997st} because it emphasizes our main observation in this section concerning the size of the 
kinetic mixing of an extra abelian interaction with the SM: 
the effective mass mixing between a new $Z'$ and the SM $Z$ boson must be rather small, as required by 
present data, but the kinetic mixing can be relatively large because it only enters alone 
in the $Z'$ interactions. 
Indeed, all deviations from the SM predictions for processes involving only SM particles and, in particular, 
testing the $Z$ boson interactions are proportional to the former and hence, it must be tiny (below the per mille level) 
because no significant departure from the SM has been observed.
While a non-vanishing kinetic mixing modifies the $Z'$ current adding an extra term to it, linear 
in the mixing and proportional to the SM hypercharge, allowing to 
constrain its size in an independent way in processes exchanging the new vector boson. 
In the case of the $Z'_{\mu - \tau}$ it means, for example, that although this gauge boson is leptophilic 
to start with, it also couples to quarks because they have non-zero hypercharge but with couplings 
suppressed by the kinetic mixing. 
The relevant question is then how small this mixing must be on phenomenological grounds. 

Let us be quantitative. Following (the notation of) Ref. \cite{Babu:1997st} the effective mass mixing, 
$\xi$, rotating the current eigenstate gauge bosons to the heavy mass eigenstate basis, can be written 
(note that we used $\xi$ before to specify the RH $Z'$ couplings in Ref. \cite{delAguila:2014soa}): 
\begin{equation}
\tan 2\xi \approx - 2 \frac{\delta M^2 + \sin \chi \ s_W \ M_Z^2}{M^2_{Z'_{\mu - \tau}}} \ ,
\label{effectivemixing}
\end{equation}
where $\delta M^2$ is the mass term mixing the gauge boson current eigenstates and $\sin \chi$ the 
corresponding kinetic mixing (and $s_W$ the sine of the EW mixing angle), 
\begin{equation}
{\cal L}_{\rm mixing} \approx \delta M^2 \ Z'^{\sigma}_{\mu - \tau} \ Z_\sigma  
- \frac{\sin \chi}{2} \ Z'^{\sigma\nu}_{\mu - \tau} \ B_{\sigma\nu} \ .
\label{mixinglagrangian}
\end{equation}
We only keep the leading 
terms in the limit of $\xi$ and $\chi$ small (fields are in the physical basis). We will concentrate on the case $\xi \ll \chi$, 
besides, which can be the result of a large mass hierarchy $M_Z^2 \ll M^2_{Z'_{\mu - \tau}}$, 
as we have formally required when 
deriving Eq. (\ref{effectivemixing}), or of a partial cancellation of the two terms in the numerator 
of this equation, or of both. A large $Z'_{\mu - \tau}$ mass can be generated by a large 
vacuum expectation value (vev) of a scalar field transforming trivially under the SM gauge symmetry; 
whereas the mass mixing cancelling the kinetic term can be obtained from a scalar multiplet 
transforming non-trivially under both the SM and the extra abelian gauge symmetry and 
developing a vev of the appropriate size. 
\footnote{Many phenomenological analyses neglect kinetic mixing effects, as, for instance, when determining 
the $ZZ'$ mixing at the LHC \cite{Andreev:2014fwa}; or the mass mixing generated by the vev of scalar fields 
transforming non-trivially under both abelian factors, as in Higgs-portal studies \cite{Choi:2013qra}. Although 
the kinetic mixing can be large and the effective mass mixing small in specific models \cite{Hirsch:2012kv}.} 

\begin{figure}
\begin{centering}
\includegraphics[width=0.49\columnwidth]{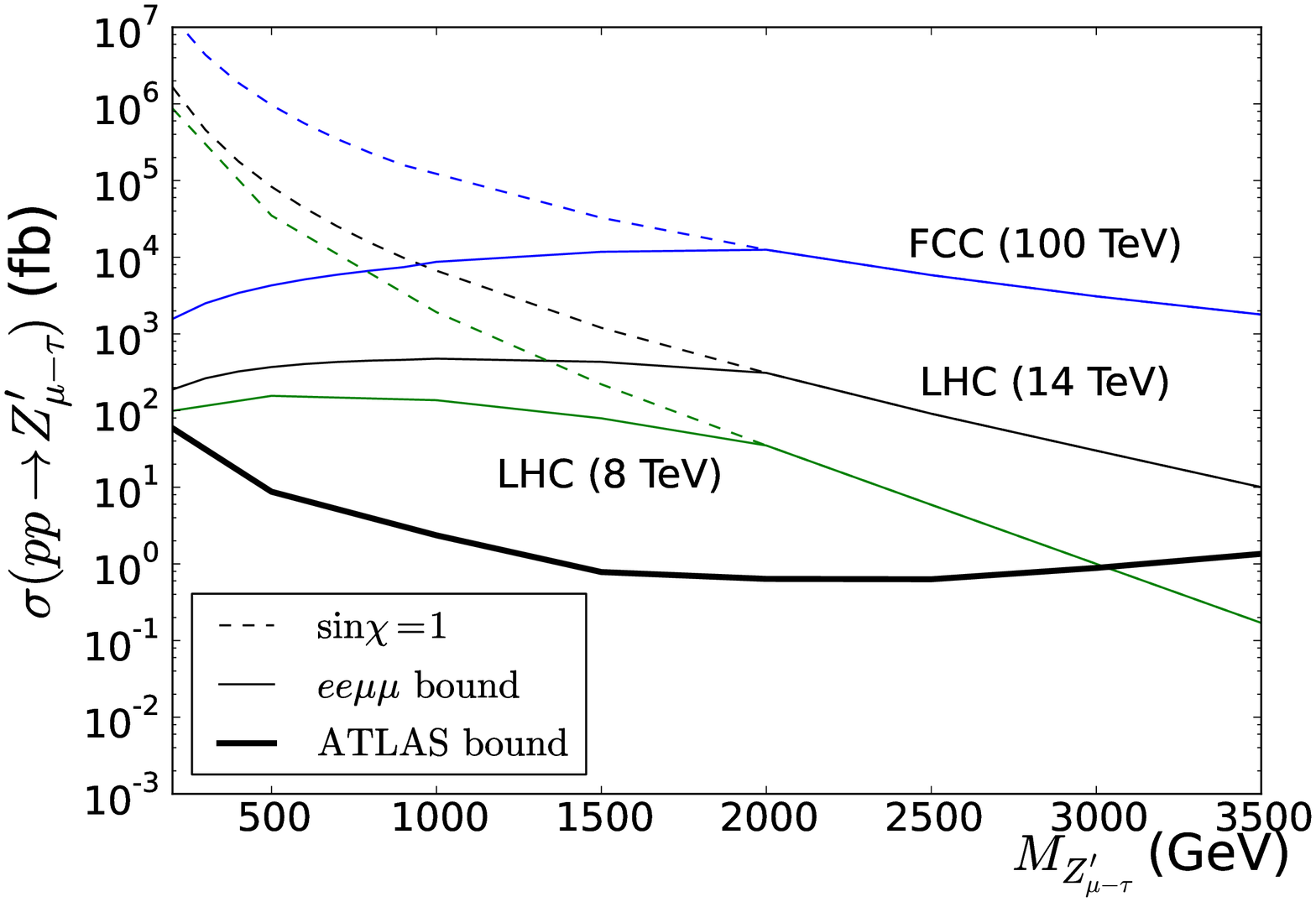}
\includegraphics[width=0.49\columnwidth]{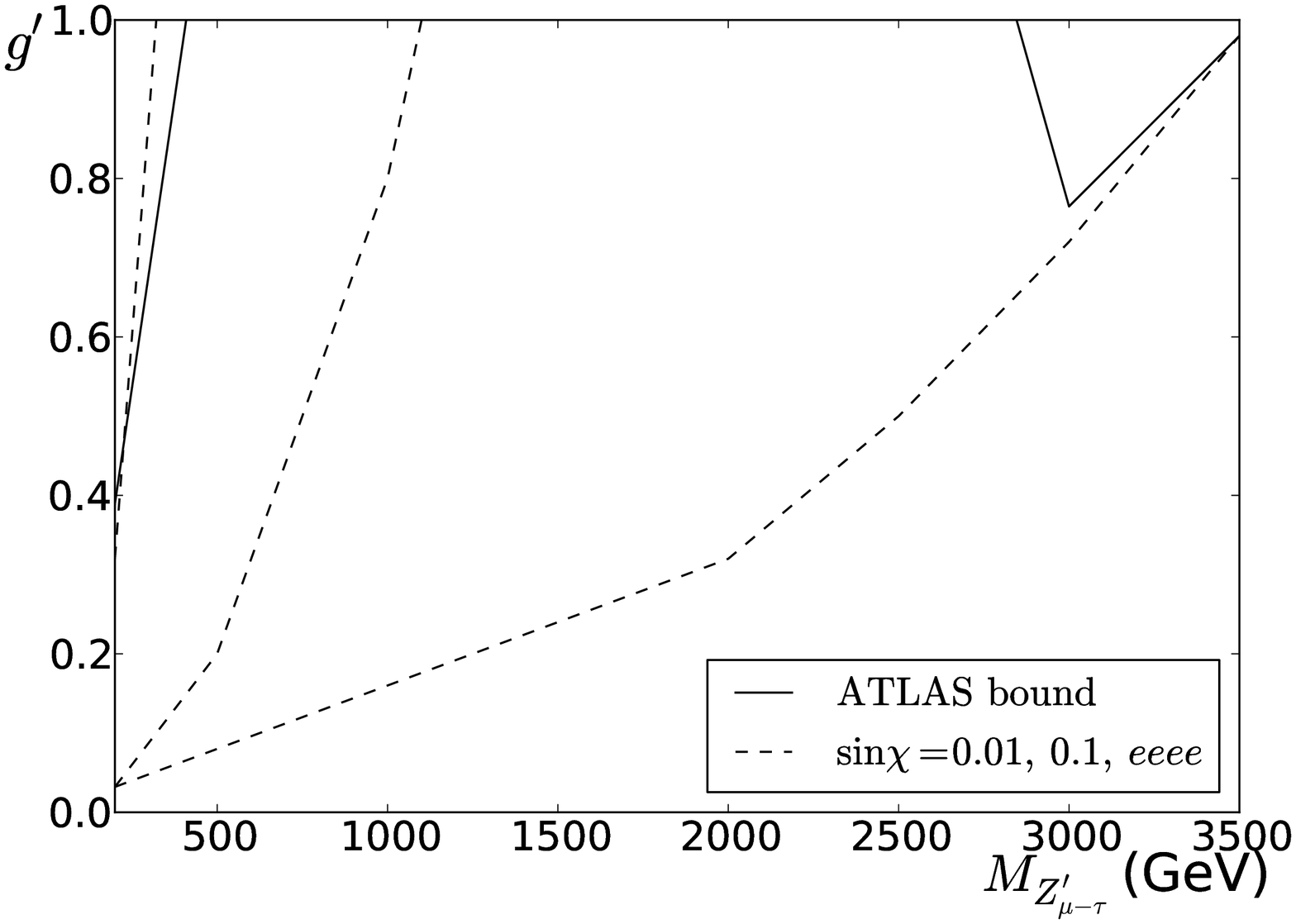}
\par\end{centering}
\caption{On the left we plot the total cross-section for $pp \rightarrow Z'_{\mu - \tau}$ 
for different values of $\sin \chi$ (see the text) as a function of the new gauge boson mass at the LHC 
and at the FCC. 
On the right we show in the $g'-M_{Z'_{\mu - \tau}}$ plane the excluded (upper) regions 
by the LEP 2 and ATLAS bounds (see the text).
\label{crosssection}
}
\end{figure}

The new gauge boson interactions with the SM fermions are described by the Lagrangian 
\begin{eqnarray}
\label{Zprimelagrangian}
{\cal L}_{Z'_{\mu - \tau}} & \approx & 
- (g' J^{Z'_{\mu - \tau}}_\sigma - \sin \chi \ \frac{e}{c_W} \ J^Y_\sigma ) \ Z'^\sigma_{\mu - \tau} \nonumber \\
& = & - [ ( g' + \sin \chi \frac{e}{2 \ c_W} )({\overline \nu_{\mu {\rm L}}} \gamma_\sigma \nu_{\mu {\rm L}} 
+ {\overline \mu_{\rm L}} \gamma_\sigma \mu_{\rm L}) + 
( g' + \sin \chi \frac{e}{c_W} ){\overline \mu_{\rm R}} \gamma_\sigma \mu_{\rm R}  
 \nonumber \\
& & + ( - g' + \sin \chi \frac{e}{2 \ c_W} )({\overline \nu_{\tau {\rm L}}} \gamma_\sigma \nu_{\tau {\rm L}} 
+ {\overline \tau_{\rm L}} \gamma_\sigma \tau_{\rm L}) + 
( - g' + \sin \chi \frac{e}{c_W} ){\overline \tau_{\rm R}} \gamma_\sigma \tau_{\rm R}  
 \nonumber \\
& & + \sin \chi \frac{e}{2 \ c_W} ({\overline \nu_{e {\rm L}}} \gamma_\sigma \nu_{e {\rm L}} 
+ {\overline e_{\rm L}} \gamma_\sigma e_{\rm L}) 
+ \sin \chi \frac{e}{c_W} {\overline e_{\rm R}} \gamma_\sigma e_{\rm R}  \\
& & - \sin \chi \frac{e}{6 \ c_W} ( {\overline u_{\rm L}} \gamma_\sigma u_{\rm L} + 
{\overline d_{\rm L}} \gamma_\sigma d_{\rm L} ) - \sin \chi \frac{2\ e}{3 \ c_W} {\overline u_{\rm R}} \gamma_\sigma u_{\rm R} 
+ \sin \chi \frac{e}{3 \ c_W} {\overline d_{\rm R}} \gamma_\sigma d_{\rm R} + \dots ] \ Z'^\sigma_{\mu - \tau} 
\nonumber \ ,
\end{eqnarray}
where the dots stand for the same quark couplings as for the first family but for the other two quark generations. 
Thus, a non-vanishing kinetic mixing modifies the $Z'_{\mu - \tau}$ current, adding an extra 
contribution linear in $\sin \chi$ and proportional to the SM hypercharge, $Y = Q - T_3$, 
with $Q$ the electric charge and $T_3$ the third component of isospin. Therefore, it 
adds $Z'_{\mu - \tau}$ couplings to electrons and to quarks, implying a new contribution 
to Drell-Yan production at the LEP 2, $e^+e^- \rightarrow Z'_{\mu - \tau} \rightarrow l\bar l$, 
and at the LHC, $q\bar q \rightarrow Z'_{\mu - \tau} \rightarrow l\bar l$. In that case, present limits on the former 
also restrict the allowed size of the kinetic mixing because no departure from the SM predictions 
has been observed but only for relatively low $Z'_{\mu - \tau}$ masses. 
In fact, the limits in Eq. (\ref{zprimebounds}) translate into (using the couplings in 
Eq. (\ref{Zprimelagrangian})):  
\begin{equation}
\frac{|\sin \chi| g'}{M^2_{Z'_{\mu - \tau}}} < 0.07 \ {\rm TeV}^{-2} \quad (ee\mu\mu) \ , \quad 
\frac{|\sin \chi|}{M_{Z'_{\mu - \tau}}} < 0.45 \ {\rm TeV}^{-1} \quad (eeee) \ .
\label{limitcoefficient}
\end{equation}
These bounds apply to the RH operators, which provide the most stringent 
limits due to the chirality dependence of the hypercharge couplings. 
The leptonic couplings in Eq. (\ref{Zprimelagrangian}) are also non-universal, 
but lepton universality was assumed in order to derive Eq. (\ref{zprimebounds}). 
In this sense, although it is a rather good approximation, the quantitative 
analysis is approximate. We also neglect higher order terms in $\sin\chi$.   
In any case, although these bounds result in small gauge couplings to quarks for small gauge boson 
masses, the direct $Z'_{\mu - \tau}$ production cross-section can be still sizable at the LHC 
if the LEP 2 limits on $\sin \chi$ are saturated.   
As a matter of fact, these bounds become unrestrictive for large vector boson masses. 
To be quantitative, we plot in Fig. \ref{crosssection} (left) the total new vector boson production 
cross-section $\sigma \ (pp \rightarrow Z'_{\mu - \tau})$, which is proportional to $|\sin \chi|^2$, 
at the LHC and the FCC for $\sin \chi = 1$ (dashed lines) and for $\sin \chi$ saturating 
the first inequality in Eq. (\ref{limitcoefficient}) with $g' = 0.25$ (solid lines). 
(Obviously, we always require $|\sin \chi| \leq 1$.) The second inequality, which is independent 
of $g'$, is less restrictive in this case, and the corresponding curve lies in between the other 
two lines plotted. 
As it is apparent, the exclusion of two-lepton cross-sections below the fb 
by the LHC and the FCC will provide quite stringent limits on the kinetic mixing 
as well as on the mass and coupling of a $Z'_{\mu - \tau}$.  
As a matter of fact, the limit reported by the ATLAS Collaboration at the LHC 
with $\sqrt s = 8$ TeV on the production cross-section of a new neutral gauge boson 
decaying into two muons 
$\sigma \ (pp \rightarrow Z') \ {\rm Br} \ (Z' \rightarrow \mu^+\mu^-)$ 
\cite{Aad:2014cka} is already more restrictive, 
for a generic $Z'_{\mu - \tau} \rightarrow \mu^+\mu^-$ branching ratio of 1/3, 
than the LEP 2 one, as shown in the same picture (lowest curve). 
It must be again emphasized that this figure is only for illustration, because 
we assumed a small enough kinetic mixing when deriving the Lagrangian in Eq. (\ref{Zprimelagrangian}), 
for instance, but draw the curves for arbitrary $\sin \chi$. 
Similarly, the gauge boson branching ratio into muons may be smaller than 1/3 for non-zero 
kinetic mixing. In fact, the ATLAS bound defines a band depending on this branching ratio 
(neglecting fermion masses): 
\begin{equation}
{\rm Br} \ (Z'_{\mu - \tau} \rightarrow \mu^+\mu^-) = \frac{1 + 1.5\ x + 0.625\ x^2 }{3 + 5\ x^2} \ , 
\quad x = \frac{\sin \chi}{g'} \frac{e}{c_W} \ ,
\label{branchingratio}
\end{equation}
which ranges from 0.01 to 0.45. 
With the same warning, on the right panel we show the excluded (upper) regions in 
the $g'-M_{Z'_{\mu - \tau}}$ plane by the first LEP 2 inequality in Eq. (\ref{limitcoefficient}) 
for $\sin \chi$ saturating the second inequality in the same equation (lowest dashed line) 
and for two other illustrative values of $\sin \chi = 0.1, 0.01$. 
The smaller the kinetic mixing is, the larger the allowed region in the $g'-M_{Z'_{\mu - \tau}}$ plane is.
We also draw the ATLAS Collaboration limit (solid line) on muon-pair production mediated by a new neutral gauge boson 
\cite{Aad:2014cka} saturating the first inequality of Eq. (\ref{limitcoefficient}), for illustration. 
Obviously, the observation of a $Z'_{\mu - \tau}$ resonance in four and two-muon events 
would allow for a determination of both parameters, $g'$ and $\sin\chi$, once fixed the gauge boson mass.

The previous discussion is based on phenomenological grounds and hence, a final comment on 
the expected size of the kinetic mixing is in order. As the studied model is renormalizable, 
quantum corrections can be actually calculated, being in general the renormalization of the kinetic 
mixing quite small and below the per mille level. 
Thus, a relatively large mixing as the one saturating Eq. (\ref{limitcoefficient}) 
requires a similar value at the matching scale. 
Therefore, although Drell-Yan production overcomes four-lepton production except for rather 
small kinetic mixing, the latter provides the most stringent model-independent limits for 
gauge boson masses below a few TeV.  

\section{Conclusions}
\label{Conclusions}

We have reviewed the LHC and ILC discovery potential for leptophilic interactions, 
which are characterized by their negligible coupling to quarks and EW bosons \cite{delAguila:2014soa}. 
In particular, we have discussed in detail the case of an extra gauge boson coupling 
to ${\rm L}_\mu - {\rm L}_\tau$, as it also provides a good estimate of the reach of these machines.  
If it only couples appreciably to muons and taus, as assumed, it can be only observed 
as a resonance in the di-muon distribution of four-lepton final states. 
Then, the corresponding Monte Carlo simulation shows that the best LHC (ILC) limits can 
be derived using 3$\mu$ plus missing energy (4$\mu$) events. 
We have found that the LHC can be sensitive up to $Z'_{\mu - \tau}$ masses near the TeV for 
$g'$ of order 1. The ILC can be only sensitive to $Z'_{\mu - \tau}$ masses below 
$300$ GeV but it could eventually provide a direct measurement of the vector boson 
couplings to tau leptons. 
Similarly, we have studied the corresponding bounds for the FCC, finding 
that it is sensitive to gauge boson masses of $\sim 2.5$ TeV for $g' \sim 1$. 
and to gauge couplings almost 5 times smaller than at the LHC for lower $Z'_{\mu - \tau}$ masses. 

All these results, however, assume a vanishing kinetic mixing of the new vector 
boson with the SM (hypercharge). Otherwise, the current coupling to the 
extra vector boson gets a new term proportional to the kinetic mixing and to 
the hypercharge and hence, the new vector boson also couples to 
electrons and to quarks but with couplings suppressed by the kinetic mixing 
and ratios fixed by the SM hypercharges. What implies new contributions 
to Drell-Yan production mediated by the new gauge boson, improving appreciably 
the LHC and ILC discovery potential due to their sensitivity to lepton pair production, 
$pp \rightarrow \gamma , Z, Z' \rightarrow l\bar l$ and 
$e^+ e^- \rightarrow \gamma , Z, Z' \rightarrow l\bar l$, 
respectively. In fact, the new contributions are in both cases proportional to the kinetic mixing square. 
Limits from LEP 2 force this mixing to be at most of few per cent for relatively 
low $Z'_{\mu - \tau}$ masses ($\sim 300$ GeV) 
but leave it essentially unconstrained for gauge boson masses of the order of few TeV. 
In contrast, the LHC data already provide the strongest bounds on this mixing, 
reaching the per mille level if no departure from the SM is observed 
with the full expected energy and luminosity. 
This limit could be eventually improved by another order of magnitude 
at the FCC.    

Few comments are in order to conclude. This study enforces the need 
to give experimental results specifying the flavor in the sampling, and 
to provide limits as model independent as possible on the corresponding 
cross-sections \cite{Aad:2014cka}. Although the kinetic mixing may receive 
small quantum corrections, its size can be large at low energy due to 
the matching with a more fundamental theory at high scale, hence making 
the phenomenological analysis compulsory. 
Finally, the SM extension with an extra gauge boson coupled 
to ${\rm L}_\mu - {\rm L}_\tau$ has been recently invoked to explain small 
discrepancies in Higgs and $B$ decays \cite{Heeck:2014qea}, although the proposed 
models include extra matter and assumptions beyond the phenomenological 
analysis performed here. 
Leptophilic interactions have been also considered to produced and annihilate 
dark matter \cite{Bell:2014tta}.
In any case, many approaches 
corner NP in the heavier families, especially when dealing with the lepton sector \cite{Alonso:2013nca}.

\section*{Acknowledgements}

This work has been supported in part by the European Commission 
through the contract PITN-GA-2012-316704 (HIGGSTOOLS), by the Ministry
of Economy and Competitiveness (MINECO), under grant numbers 
FPA2010-17915 and FPA2013-47836-C3-1,2-P, and by the Junta de Andaluc{\'\i}a 
grants FQM 101 and FQM 6552.


\begin{thebibliography}{99}

%%%%% SM Higgs mechanism %%%%%

%\cite{Englert:1964et}
\bibitem{Englert:1964et}
  F.~Englert and R.~Brout,
  %``Broken Symmetry and the Mass of Gauge Vector Mesons,''
  Phys.\ Rev.\ Lett.\  {\bf 13} (1964) 321;
  %%CITATION = PRLTA,13,321;%%
  %2608 citations counted in INSPIRE as of 06 Oct 2014
%\cite{Higgs:1964pj}
%\bibitem{Higgs:1964pj}
  P.~W.~Higgs,
  %``Broken Symmetries and the Masses of Gauge Bosons,''
  Phys.\ Rev.\ Lett.\  {\bf 13} (1964) 508.
  %%CITATION = PRLTA,13,508;%%
  %2856 citations counted in INSPIRE as of 06 Oct 2014

%%%%% SM Higgs discovery %%%%%

%\cite{Aad:2012tfa}
\bibitem{Aad:2012tfa}
  G.~Aad {\it et al.}  [ATLAS Collaboration],
  %``Observation of a new particle in the search for the Standard Model Higgs boson with the ATLAS detector at the LHC,''
  Phys.\ Lett.\ B {\bf 716} (2012) 1
  [arXiv:1207.7214 [hep-ex]];
  %%CITATION = ARXIV:1207.7214;%%
  %908 citations counted in INSPIRE as of 11 Apr 2013
%\cite{Chatrchyan:2012ufa}
%\bibitem{Chatrchyan:2012ufa}
  S.~Chatrchyan {\it et al.}  [CMS Collaboration],
  %``Observation of a new boson at a mass of 125 GeV with the CMS experiment at the LHC,''
  Phys.\ Lett.\ B {\bf 716} (2012) 30
  [arXiv:1207.7235 [hep-ex]].
  %%CITATION = ARXIV:1207.7235;%%
  %899 citations counted in INSPIRE as of 11 Apr 2013

%%%%% Examples:  %%%%% 

%\cite{Barcelo:2011vk}
\bibitem{Barcelo:2011vk}
  R.~Barcelo, A.~Carmona, M.~Masip and J.~Santiago,
  %``Stealth gluons at hadron colliders,''
  Phys.\ Lett.\ B {\bf 707} (2012) 88
  [arXiv:1106.4054 [hep-ph]];
  %%CITATION = ARXIV:1106.4054;%%
  %57 citations counted in INSPIRE as of 24 Apr 2015
see for related work and further references: 
  %\cite{Aguilar-Saavedra:2014kpa}
%\bibitem{Aguilar-Saavedra:2014kpa}
  J.~A.~Aguilar-Saavedra, D.~Amidei, A.~Juste and M.~Perez-Victoria,
  %``Asymmetries in top quark pair production at hadron colliders,''
  arXiv:1406.1798 [hep-ph].
  %%CITATION = ARXIV:1406.1798;%%
  %14 citations counted in INSPIRE as of 24 Apr 2015  

%\cite{Farrar:1978xj}
\bibitem{Farrar:1978xj}
  G.~R.~Farrar and P.~Fayet,
  %``Phenomenology of the Production, Decay, and Detection of New Hadronic States Associated with Supersymmetry,''
  Phys.\ Lett.\ B {\bf 76} (1978) 575;  
  %%CITATION = PHLTA,B76,575;%%
  %1059 citations counted in INSPIRE as of 24 Apr 2015
see for a recent review and further references: 
%\cite{Mohapatra:2015fua}
%\bibitem{Mohapatra:2015fua}
  R.~N.~Mohapatra,
  %``Supersymmetry and R-parity: an Overview,''
  arXiv:1503.06478 [hep-ph].
  %%CITATION = ARXIV:1503.06478;%%
  %1 citations counted in INSPIRE as of 24 Apr 2015  
  
%%%%% ILC %%%%% 

%\cite{Baer:2013cma}
\bibitem{Baer:2013cma}
  H.~Baer, T.~Barklow, K.~Fujii, Y.~Gao, A.~Hoang, S.~Kanemura, J.~List and H.~E.~Logan {\it et al.},
  %``The International Linear Collider Technical Design Report - Volume 2: Physics,''
  arXiv:1306.6352 [hep-ph], and references there in.
  %%CITATION = ARXIV:1306.6352;%%
  %48 citations counted in INSPIRE as of 07 Jan 2014

%%%%% Leptophilic interactions %%%%% 

%\cite{delAguila:2014soa}
\bibitem{delAguila:2014soa}
  F.~del Aguila, M.~Chala, J.~Santiago and Y.~Yamamoto,
  %``Collider limits on leptophilic interactions,''
  JHEP {\bf 1503} (2015) 059
  [arXiv:1411.7394 [hep-ph]].
  %%CITATION = ARXIV:1411.7394;%%
  %3 citations counted in INSPIRE as of 31 Mar 2015

%%%%%%%%%%% Zmu-tau %%%%%%%%%%%% 

%\cite{Foot:1994vd}
\bibitem{Foot:1994vd}
  R.~Foot, X.~G.~He, H.~Lew and R.~R.~Volkas,
  %``Model for a light Z-prime boson,''
  Phys.\ Rev.\ D {\bf 50} (1994) 4571
  [hep-ph/9401250]; and references there in.
  %%CITATION = HEP-PH/9401250;%%
  %24 citations counted in INSPIRE as of 27 Dec 2013
    
%%%%% FCC %%%%% 

\bibitem{hadronFCC}
A. Ball et al., {\it Future Circular Collider Study, Hadron Collider Parameters}, Tech. Rep. FCC-ACC-SPC-0001, CERN, Geneva, 2014.

\bibitem{leptonFCC}
J. Wenninger et al., {\it Future Circular Collider Study, Lepton Collider Parameters}, Tech. Rep. FCC-ACC-SPC-0003 (Rev. 2.0), CERN, Geneva, 2014. 

%\cite{AbelleiraFernandez:2012cc}
\bibitem{AbelleiraFernandez:2012cc}
  J.~L.~Abelleira Fernandez {\it et al.}  [LHeC Study Group Collaboration],
  %``A Large Hadron Electron Collider at CERN: Report on the Physics and Design Concepts for Machine and Detector,''
  J.\ Phys.\ G {\bf 39} (2012) 075001
  [arXiv:1206.2913 [physics.acc-ph]].
  %%CITATION = ARXIV:1206.2913;%%
  %153 citations counted in INSPIRE as of 24 Apr 2015
    
%%%%%%%%%%% large Z' kinetic mixing %%%%%%%%%%%% 

%\cite{Babu:1997st}
\bibitem{Babu:1997st}
  K.~S.~Babu, C.~F.~Kolda and J.~March-Russell,
  %``Implications of generalized Z - Z-prime mixing,''
  Phys.\ Rev.\ D {\bf 57} (1998) 6788
  [hep-ph/9710441].
  %%CITATION = HEP-PH/9710441;%%
  %150 citations counted in INSPIRE as of 31 Mar 2015
  
%%%%% LNV scalar production %%%%% 

%\cite{delAguila:2013yaa}
\bibitem{delAguila:2013yaa}
 F.~del Aguila, M.~Chala, A.~Santamaria and J.~Wudka,
  %``Discriminating between lepton number violating scalars using events with four and three charged leptons at the LHC,''
  Phys.\ Lett.\ B {\bf 725} (2013) 310
  [arXiv:1305.3904 [hep-ph]]; 
  %%CITATION = ARXIV:1305.3904;%%
  %10 citations counted in INSPIRE as of 12 Nov 2014
%\cite{delAguila:2013hla}
%\bibitem{delAguila:2013hla}
%  F.~del Aguila, M.~Chala, A.~Santamaria and J.~Wudka,
  %``Distinguishing between lepton number violating scalars at the LHC,''
  EPJ Web Conf.\  {\bf 60} (2013) 17002
  [arXiv:1307.0510 [hep-ph]]; 
  %%CITATION = ARXIV:1307.0510;%%
  %5 citations counted in INSPIRE as of 12 Nov 2014
%\cite{delAguila:2013mia}
%\bibitem{delAguila:2013mia}
  F.~del Aguila and M.~Chala,
  %``LHC bounds on Lepton Number Violation mediated by doubly and singly-charged scalars,''
  JHEP {\bf 1403} (2014) 027
  [arXiv:1311.1510 [hep-ph]]; 
  %%CITATION = ARXIV:1311.1510;%%
  %10 citations counted in INSPIRE as of 12 Nov 2014
 %\cite{delAguila:2013aga}
%\bibitem{delAguila:2013aga}
  F.~del Aguila, M.~Chala, A.~Santamaria and J.~Wudka,
  %``Lepton Number Violation and Scalar Searches at the LHC,''
  Acta Phys.\ Polon.\ B {\bf 44} (2013) 11,  2139
  [arXiv:1311.2950 [hep-ph]].
  %%CITATION = ARXIV:1311.2950;%%
  %1 citations counted in INSPIRE as of 12 Nov 2014

%%%%%%%%%%% tau custodians %%%%%%%%%%%% 

%\cite{delAguila:2010vg}
\bibitem{delAguila:2010vg}
  F.~del Aguila, A.~Carmona and J.~Santiago,
  %``Neutrino Masses from an A4 Symmetry in Holographic Composite Higgs Models,''
  JHEP {\bf 1008} (2010) 127
  [arXiv:1001.5151 [hep-ph]]; 
  %%CITATION = ARXIV:1001.5151;%%
  %52 citations counted in INSPIRE as of 11 Apr 2015
%\cite{delAguila:2010es}
%\bibitem{delAguila:2010es}
%  F.~del Aguila, A.~Carmona and J.~Santiago,
  %``Tau Custodian searches at the LHC,''
  Phys.\ Lett.\ B {\bf 695} (2011) 449
  [arXiv:1007.4206 [hep-ph]].
  %%CITATION = ARXIV:1007.4206;%%
  %29 citations counted in INSPIRE as of 11 Apr 2015
  
 %%%%% LEP and global bounds %%%%% 
   
 %\cite{Schael:2013ita}
\bibitem{Schael:2013ita}
  S.~Schael {\it et al.}  [ALEPH and DELPHI and L3 and OPAL and LEP Electroweak Collaborations],
  %``Electroweak Measurements in Electron-Positron Collisions at W-Boson-Pair Energies at LEP,''
  Phys.\ Rept.\  {\bf 532} (2013) 119
  [arXiv:1302.3415 [hep-ex]].
  %%CITATION = ARXIV:1302.3415;%%
  %67 citations counted in INSPIRE as of 20 Nov 2014 
  
%\cite{Blas:2013ana}
\bibitem{Blas:2013ana}
  J.~de Blas,
  %``Electroweak limits on physics beyond the Standard Model,''
  EPJ Web Conf.\  {\bf 60} (2013) 19008
  [arXiv:1307.6173 [hep-ph]].
  %%CITATION = ARXIV:1307.6173;%%
  %2 citations counted in INSPIRE as of 03 Apr 2014    

  %\cite{delAguila:2010mx}
\bibitem{delAguila:2010mx}
  F.~del Aguila, J.~de Blas and M.~Perez-Victoria,
  %``Electroweak Limits on General New Vector Bosons,''
  JHEP {\bf 1009} (2010) 033
  [arXiv:1005.3998 [hep-ph]]; 
  %%CITATION = ARXIV:1005.3998;%%
  %55 citations counted in INSPIRE as of 14 Feb 2014
%\cite{delAguila:2011zs}
%\bibitem{delAguila:2011zs}
  F.~del Aguila and J.~de Blas,
  %``Electroweak constraints on new physics,''
  Fortsch.\ Phys.\  {\bf 59} (2011) 1036
  [arXiv:1105.6103 [hep-ph]]; 
  %%CITATION = ARXIV:1105.6103;%%
  %10 citations counted in INSPIRE as of 12 Nov 2014  
%\cite{deBlas:2013qqa}
%\bibitem{deBlas:2013qqa}
  J.~de Blas, M.~Chala and J.~Santiago,
  %``Global Constraints on Lepton-Quark Contact Interactions,''
  Phys.\ Rev.\ D {\bf 88} (2013) 095011
  [arXiv:1307.5068 [hep-ph]].
  %%CITATION = ARXIV:1307.5068;%%
  %4 citations counted in INSPIRE as of 03 Apr 2014
  
%%%%% Neutrino trident production %%%%% 

%\cite{Altmannshofer:2014cfa}
\bibitem{Altmannshofer:2014cfa}
  W.~Altmannshofer, S.~Gori, M.~Pospelov and I.~Yavin,
  %``Dressing L_mu - L_tau in Color,''
  Phys.\ Rev.\ D {\bf 89} (2014) 095033
  [arXiv:1403.1269 [hep-ph]]; 
  %%CITATION = ARXIV:1403.1269;%%
  %10 citations counted in INSPIRE as of 02 Sep 2014
%\cite{Altmannshofer:2014pba}
%\bibitem{Altmannshofer:2014pba}
%  W.~Altmannshofer, S.~Gori, M.~Pospelov and I.~Yavin,
  %``Neutrino Trident Production: A Powerful Probe of New Physics with Neutrino Beams,''
  Phys.\ Rev.\ Lett.\  {\bf 113} (2014) 091801
  [arXiv:1406.2332 [hep-ph]].
  %%CITATION = ARXIV:1406.2332;%%
  %13 citations counted in INSPIRE as of 12 Apr 2015

%%%%% ZZ' mixing %%%%% 

%\cite{Holdom:1985ag}
\bibitem{Holdom:1985ag}
  B.~Holdom,
  %``Two U(1)'s and Epsilon Charge Shifts,''
  Phys.\ Lett.\ B {\bf 166} (1986) 196.
  %%CITATION = PHLTA,B166,196;%%
  %743 citations counted in INSPIRE as of 22 Sep 2014

%\cite{del Aguila:1985cb}
\bibitem{del Aguila:1985cb}
  F.~del Aguila, G.~A.~Blair, M.~Daniel and G.~G.~Ross,
  %``Superstring Inspired Models,''
  Nucl.\ Phys.\ B {\bf 272} (1986) 413; 
  %%CITATION = NUPHA,B272,413;%%
  %124 citations counted in INSPIRE as of 22 Sep 2014
%\cite{delAguila:1988jz}
%\bibitem{delAguila:1988jz}
  F.~del Aguila, G.~D.~Coughlan and M.~Quiros,
  %``Gauge Coupling Renormalization With Several U(1) Factors,''
  Nucl.\ Phys.\ B {\bf 307} (1988) 633
   [Erratum-ibid.\ B {\bf 312} (1989) 751];
  %%CITATION = NUPHA,B307,633;%%
  %89 citations counted in INSPIRE as of 22 Sep 2014
%\cite{Foot:1991kb}
%\bibitem{Foot:1991kb}
  R.~Foot and X.~G.~He,
  %``Comment on Z Z-prime mixing in extended gauge theories,''
  Phys.\ Lett.\ B {\bf 267} (1991) 509; 
  %%CITATION = PHLTA,B267,509;%%
  %134 citations counted in INSPIRE as of 24 Apr 2015
%\cite{Carone:1995pu}
%\bibitem{Carone:1995pu}
  C.~D.~Carone and H.~Murayama,
  %``Realistic models with a light U(1) gauge boson coupled to baryon number,''
  Phys.\ Rev.\ D {\bf 52} (1995) 484
  [hep-ph/9501220];
  %%CITATION = HEP-PH/9501220;%%
  %53 citations counted in INSPIRE as of 22 Sep 2014
%\cite{delAguila:1995rb}
%\bibitem{delAguila:1995rb}
  F.~del Aguila, M.~Masip and M.~Perez-Victoria,
  %``Physical parameters and renormalization of U(1)-a x U(1)-b models,''
  Nucl.\ Phys.\ B {\bf 456} (1995) 531
  [hep-ph/9507455]; 
  %%CITATION = HEP-PH/9507455;%%
  %52 citations counted in INSPIRE as of 22 Sep 2014
see for a review and further references: 
%\cite{Langacker:2008yv}
%\bibitem{Langacker:2008yv}
  P.~Langacker,
  %``The Physics of Heavy $Z^\prime$ Gauge Bosons,''
  Rev.\ Mod.\ Phys.\  {\bf 81} (2009) 1199
  [arXiv:0801.1345 [hep-ph]]; 
  %%CITATION = ARXIV:0801.1345;%%
  %597 citations counted in INSPIRE as of 24 Apr 2015
%\cite{Agashe:2014kda}
%\bibitem{Agashe:2014kda}
  K.~A.~Olive {\it et al.}  [Particle Data Group Collaboration],
  %``Review of Particle Physics,''
  Chin.\ Phys.\ C {\bf 38} (2014) 090001. 
  %%CITATION = CHPHD,C38,090001;%%
  %943 citations counted in INSPIRE as of 24 Apr 2015

%%% No kinetic mixing %%%

  %\cite{Andreev:2014fwa}
\bibitem{Andreev:2014fwa}
  V.~V.~Andreev, P.~Osland and A.~A.~Pankov,
  %``Precise determination of Z-Z' mixing at the CERN LHC,''
  Phys.\ Rev.\ D {\bf 90} (2014) 055025
  [arXiv:1406.6776 [hep-ph]].
  %%CITATION = ARXIV:1406.6776;%%
  %2 citations counted in INSPIRE as of 18 Nov 2014

%%% No mass mixing %%%

 %\cite{Choi:2013qra}
\bibitem{Choi:2013qra}
  S.~Y.~Choi, C.~Englert and P.~M.~Zerwas,
  %``Multiple Higgs-Portal and Gauge-Kinetic Mixings,''
  Eur.\ Phys.\ J.\ C {\bf 73} (2013) 2643
  [arXiv:1308.5784 [hep-ph]].
  %%CITATION = ARXIV:1308.5784;%%
  %12 citations counted in INSPIRE as of 24 Apr 2015 
 
%%% Large kinetic mixing and small effective mass mixing %%%

%\cite{Hirsch:2012kv}
\bibitem{Hirsch:2012kv}
  M.~Hirsch, W.~Porod, L.~Reichert and F.~Staub,
  %``Phenomenology of the minimal supersymmetric $U(1)_{B-L}\times U(1)_R$ extension of the standard model,''
  Phys.\ Rev.\ D {\bf 86} (2012) 093018
  [arXiv:1206.3516 [hep-ph]]; 
  %%CITATION = ARXIV:1206.3516;%%
  %43 citations counted in INSPIRE as of 24 Apr 2015
see also: 
%\cite{DelAguila:1993px}
%\bibitem{DelAguila:1993px}
  F.~del Aguila,
  %``The Physics of z-prime bosons,''
  Acta Phys.\ Polon.\ B {\bf 25} (1994) 1317
  [hep-ph/9404323].
  %%CITATION = HEP-PH/9404323;%%
  %34 citations counted in INSPIRE as of 24 Apr 2015 
 
%\cite{Aad:2014cka}
\bibitem{Aad:2014cka}
  G.~Aad {\it et al.}  [ATLAS Collaboration],
  %``Search for high-mass dilepton resonances in pp collisions at $\sqrt{s}=8$??TeV with the ATLAS detector,''
  Phys.\ Rev.\ D {\bf 90} (2014) 5,  052005
  [arXiv:1405.4123 [hep-ex]]; 
  %%CITATION = ARXIV:1405.4123;%%
  %58 citations counted in INSPIRE as of 23 Apr 2015
%\cite{Aad:2015osa}
%\bibitem{Aad:2015osa}
%  G.~Aad {\it et al.}  [ATLAS Collaboration],
  %``A search for high-mass resonances decaying to $\tau^{+}\tau^{-}$ in $pp$ collisions at $\sqrt{s}=8$ TeV with the ATLAS detector,''
  arXiv:1502.07177 [hep-ex].
  %%CITATION = ARXIV:1502.07177;%%

%%%%%%%%%%% Zmu-tau effects and limits %%%%%%%%%%%% 
 
%\cite{Heeck:2014qea}
\bibitem{Heeck:2014qea}
  J.~Heeck, M.~Holthausen, W.~Rodejohann and Y.~Shimizu,
  %``Higgs $\to \mu \tau$ in Abelian and Non-Abelian Flavor Symmetry Models,''
  arXiv:1412.3671 [hep-ph]; 
  %%CITATION = ARXIV:1412.3671;%%
  %11 citations counted in INSPIRE as of 24 Apr 2015  
%\cite{Crivellin:2015mga}
%\bibitem{Crivellin:2015mga}
  A.~Crivellin, G.~D'Ambrosio and J.~Heeck,
  %``Explaining $h\to\mu^\pm\tau^\mp$, $B\to K^* \mu^+\mu^-$ and $B\to K \mu^+\mu^-/B\to K e^+e^-$ in a two-Higgs-doublet model with gauged $L_\mu-L_\tau$,''
  arXiv:1501.00993 [hep-ph]; 
  %%CITATION = ARXIV:1501.00993;%%
  %5 citations counted in INSPIRE as of 11 Mar 2015 
 %\cite{Crivellin:2015lwa}
%\bibitem{Crivellin:2015lwa}
%  A.~Crivellin, G.~DÕAmbrosio and J.~Heeck,
  %``Addressing the LHC flavor anomalies with horizontal gauge symmetries,''
  Phys.\ Rev.\ D {\bf 91} (2015) 7,  075006
  [arXiv:1503.03477 [hep-ph]].
  %%CITATION = ARXIV:1503.03477;%%
  %6 citations counted in INSPIRE as of 24 Apr 2015 
  
  %\cite{Bell:2014tta}
\bibitem{Bell:2014tta}
  N.~F.~Bell, Y.~Cai, R.~K.~Leane and A.~D.~Medina,
  %``Leptophilic dark matter with $Z?$ interactions,''
  Phys.\ Rev.\ D {\bf 90} (2014) 3,  035027
  [arXiv:1407.3001 [hep-ph]].
  %%CITATION = ARXIV:1407.3001;%%
  %9 citations counted in INSPIRE as of 24 Apr 2015 
  
 %\cite{Alonso:2013nca}
\bibitem{Alonso:2013nca}
  R.~Alonso, M.~B.~Gavela, G.~Isidori and L.~Maiani,
  %``Neutrino Mixing and Masses from a Minimum Principle,''
  JHEP {\bf 1311} (2013) 187
  [arXiv:1306.5927 [hep-ph]].
  %%CITATION = ARXIV:1306.5927;%%
  %17 citations counted in INSPIRE as of 23 Apr 2015

\end{thebibliography}
\end{document}